\RequirePackage{ifpdf}

\documentclass[hyper,letterpaper]{JHEP3}
\usepackage{amsmath,amssymb,amsfonts,amsthm}
\usepackage{fancybox}
\usepackage{cite}
\usepackage{graphicx, wrapfig}
\usepackage{multirow}
\usepackage{verbatim}
\usepackage{appendix}
\usepackage{slashed}
\usepackage{url}
\usepackage{float}
\DeclareRobustCommand{\quantumbinomial}{\genfrac{[}{]}{0pt}{}}
\newtheorem*{conj}{Conjecture}
\allowdisplaybreaks

\title{Entanglement entropy from SU(2) Chern-Simons theory and symmetric webs}

\author{Sungbong Chun$^{1}$, Ning Bao$^{1,2}$
\\
$^{1}$Walter Burke Institute for Theoretical Physics, California Institute of Technology, Pasadena, CA 91125 USA\\
$^{2}$Institute of Quantum Information and Matter, California Institute of Technology, Pasadena, CA 91125, USA}

\abstract{A path integral on a link complement of a three-sphere fixes a vector (the "link state") in Chern-Simons theory. The link state can be written in a certain basis with the colored link invariants as its coefficients. We use symmetric webs to systematically compute the colored link invariants, by which we can write down the multi-partite entangled state of any given link. It is still unknown if a product state necessarily implies that the corresponding components are unlinked, and we leave it as a conjecture. 
\\
\\
\\
\\
\\
{\tt CALT-TH-2017-35}}

\begin{document}
\cornersize{1}

\section{Introduction}
\label{sec:intro}
The study of entanglement entropy has been a deeply fascinating and instructive one in both the fields of high energy \cite{Ryu:2006bv} and condensed matter physics \cite{Laflorencie:2015eck}. In particular, it appears to have deep connections with both geometric and topological properties in both disciplines \cite{Ryu:2006bv, 2007JSMTE..08...24H, Kitaev:2005dm}. Here we are interested in extending this study in the specific realm of Chern-Simons gauge theory using topological techniques.

\begin{figure} [htb]
\centering
\includegraphics{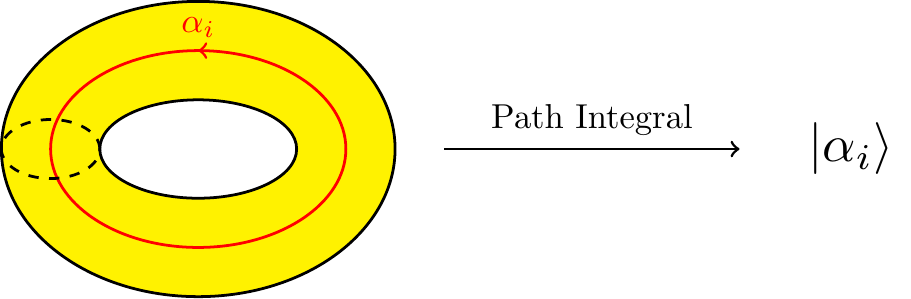}
\caption{A solid torus (colored in yellow) contains a Wilson loop (colored in red), which wraps its non-contractible cycle. When the Wilson loop carries a representation $\alpha_{i}$ of the gauge group, the path integral fixes a vector $|\alpha_{i} \rangle$ in $H_{T^{2}}$.}
\label{fig:solidtorus}
\end{figure}

Chern-Simons gauge theory is a three-dimensional TQFT with non-local gauge invariant observables, the Wilson loop operators \cite{Witten89}. Performing a path integral on a 3-manifold with boundary, one obtains a vector in the space of conformal blocks of the associated WZW model on the boundary 2-manifold(s). In particular, a path integral on the link complement determines a ``link state'':

$$| \mathcal{L} \rangle = \sum_{\alpha_{1}, \cdots, \alpha_{m}} C(\alpha_{1}, \cdots, \alpha_{m}) | \alpha_{1} \rangle \otimes  \cdots \otimes | \alpha_{m} \rangle,$$
where $\mathcal{L}$ is a $m$-component link, and $\alpha_{i}$'s are integrable representations of the gauge group. Each vector $| \alpha_{i} \rangle $ belongs to the 2d Hilbert space associated to a torus $H_{T^{2}}$, which is fixed by a path integral on a solid torus with a Wilson loop colored in $\alpha_{i}$, as in Figure \ref{fig:solidtorus}. The inner product of the Hilbert space $H_{T^{2}}$ is simply $\langle \alpha_{i} | \alpha_{j} \rangle = \delta_{\alpha_{i},\alpha_{j}}$, and thus, $(\langle \alpha_{1} | \otimes \cdots \otimes \langle \alpha_{m}| ) | \mathcal{L} \rangle = C(\alpha_{1},\cdots,\alpha_{m})$. Topologically, $(\langle \alpha_{1} | \otimes \cdots \otimes \langle \alpha_{m}| ) | \mathcal{L} \rangle$ stands for gluing $m$ solid tori with Wilson loops colored in $\alpha_{i}$ back into the link complement, so $C(\alpha_{1},\cdots,\alpha_{m})$ is nothing but a ``colored'' link invariant of $\mathcal{L}$. Once these colored link invariants of $\mathcal{L}$ are known for all possible colorings, one can compute the density matrix and the entanglement entropy of $\mathcal{L}$ and check whether the link state is an entangled state or not \cite{BFLP,Salton:2016qpp}. 

The aim of our paper is to reinforce \cite{BFLP,Salton:2016qpp} when the gauge group is $SU(2)$ with a technique to compute the colored link invariants $C(\alpha_{1}, \cdots, \alpha_{m})$ for any link $\mathcal{L}$, for any given coloring. To clarify, \cite{BFLP,Salton:2016qpp} computes the link states and study their entanglement properties when the given links are torus links (so that their colored link invariants can be written as a product of $S$ and $T$ matrices of the associated WZW model) or twist links (in which case Habiro's formula works nicely). In this paper, we use junctions of Wilson lines and the ``symmetric web'' relations \cite{RoseTub}, by which one can systematically compute the colored invariants for more general classes of knots/links. This will allow for the construction of toplogically interesting states and sets of entanglement entropies via topological techniques. One step further, we propose a conjecture in which the entanglement data is transcribed to the topological data: namely, if the link state factorizes for all colorings $\alpha_{i}$'s, then the corresponding link itself  is reducible, \textit{i.e.}, a union of two unlinked sub-links. In terms of colored Jones polynomials, we can phrase the conjecture as follows:

\begin{conj}
Given a $m$-component link $\mathcal{L}$, suppose there exist two sub-links $\mathcal{L}_{1}$ and $\mathcal{L}_{2}$, each with $i$ and $(m-i)$ components. Suppose the two sub-links satisfy the following:
$$J_{\alpha_{1}, \cdots, \alpha_{m}}(\mathcal{L}) = J_{\alpha_{1}, \cdots, \alpha_{i}}(\mathcal{L}_{1})J_{\alpha_{i+1}, \cdots, \alpha_{m}}(\mathcal{L}_{2})$$ 
for all colorings $\alpha_{1}, \cdots, \alpha_{m}$, then $\mathcal{L}_{1}$ and $\mathcal{L}_{2}$ are unlinked.
\end{conj} 

\section{Review: the link states and entanglement entropy}
First, let's recall the  definition of entanglement entropy. Recall that the entanglement entropy $S$ of a system $A$ is given by

$$S_A=-\mathrm{Tr} \rho_A \log \rho_A,$$

where $\rho_A$ is the reduced density matrix corresponding to system $A$. The entanglement entropy quantifies the amount of entanglement, whether it be classical correlation or genuine quantum entanglement, that $A$ shares with all other systems. Note that $S_A$ is zero for pure states and nonzero for mixed states. In general, the calculation of the entanglement entropy is somewhat involved, particularly for large Hilbert spaces for which taking the logarithm of a high-dimensional matrix is computationally expensive.

Next, let us briefly review \cite{Witten89}. We start with $SU(2)$ Chern-Simons theory at level $k$ on a 3-manifold $M_{3}$:

$$S_{CS} = \frac{k}{4\pi} \mathrm{Tr} \int_{M_{3}} A \wedge dA + \frac{2}{3}A \wedge A \wedge A,$$
where $A$ is a $su(2)$-valued one form on $M_{3}$. When $M_{3}$ is a closed 3-manifold, the partition function $Z(M_{3}) = \int [\mathcal{D}A] e^{i S_{CS}}$ defines a topological 3-manifold invariant. We can also introduce gauge invariant observables, the famous Wilson loops defined on a closed path $C$ in $M_{3}$:
$$W_{R}(C) = \mathrm{Tr}_{R} \bigg[ \mathcal{P} \oint_{C} e^{i A} \bigg],$$
where $R$ is an irreducible representation of $SU(2)$. When $M_{3}$ is a 3-sphere and $R = \square$ (the fundamental representation of $SU(2)$), the expectation value of Wilson loops coincides with the Jones polynomials \cite{Witten89}:
\begin{equation}
\langle W_{\square}(C) \rangle = \dfrac{\int [\mathcal{D}A] W_{\square}(C) e^{i S_{CS}}}{\int [\mathcal{D}A] e^{i S_{CS}}} = J_{\square}(C).
\label{eqn:Jones}
\end{equation}
Equation \ref{eqn:Jones} holds, because $\langle W_{\square}(unknot) \rangle$ equals $J_{\square}(unknot)$ and the skein relation holds. Indeed, gluing two solid tori via modular S-transform on the boundary, one obtains a 3-sphere. When only one of the tori has a Wilson loop colored in $\square$, path integral on these solid tori fixes vctors $|\square \rangle$ and $|0 \rangle$ in the associated Hilbert space $H_{T^{2}}$. Then, the expectation value of the Wilson loop can be written as: 

$$\langle W_{\square}(unknot) \rangle = \frac{\langle 0 | S | \square \rangle}{\langle 0 | S |  0 \rangle } = \frac{S_{0\square}}{S_{00}} = J_{\square}(unknot).$$
where $S_{ij}$ is the S-matrix element of $\widehat{su(2)}_{k}$ WZW model on torus. Here, $i, j$ stand for the integrable representations of the affine Lie algebra $\widehat{su(2)}_{k}$, which are in one-to-one correspondence with the space of conformal blocks $H_{T^{2}}$. In case the gauge group is $SU(2)$ and the level is $k$, the integrable representations are spin-$j$ representations, where $j = 0, 1/2, \cdots, k/2$. Thus, $H_{T^{2}}$ is spanned by the vectors $|0 \rangle, |\tfrac{1}{2}\rangle, \cdots, |\tfrac{k}{2} \rangle$, and $|\square\rangle = |\tfrac{1}{2}\rangle$ in our notation.

\begin{figure} [htb]
\centering
\includegraphics{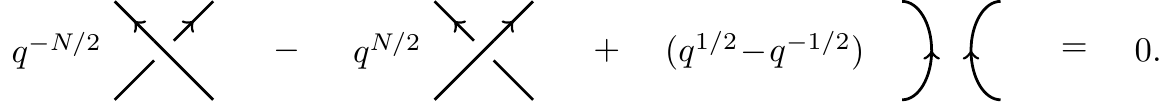}
\caption{The skein relation among three Wilson lines in $D^{3}$. $N$ is the rank of the gauge group, and all three Wilson lines are in the fundamental representation ($\square$) and canonically framed.}
\label{fig:skein}
\end{figure}
Next, consider the local relation among three Wilson lines shown in Figure \ref{fig:skein}. The three Wilson lines are lying in a closed three-ball $D^{3}$, and they are related to each other by half-twist(s) along the vertical direction. Performing a path integral on $D^{3}$ containing any one of the above three Wilson lines would fix a vector in the Hilbert space $H_{S^{2};\square,\square,\bar{\square},\bar{\square}}$ associated to the punctured boundary 2-sphere. The Hilbert space is 2-dimensional by the charge conservation argument, so the three Wilson lines shown in Figure \ref{fig:skein} must satisfy a linear relation in this 2d Hilbert space. The coefficients are fixed by studying the action of the half-twist on $S^{2}$. Since the Wilson lines colored in $\square$ satisfy $\langle W_{\square}(unknot) \rangle = J_{\square}(unknot)$ and the skein relation, we may repeatedly apply the skein relation until no crossing remains. Then, $\langle W_{\square}(unknot) \rangle = J_{\square}(unknot)$ determines the expectation value of the original Wilson loop $C$, and Equation \ref{eqn:Jones} holds.

The space $H_{T^{2}}$ is also equipped with a metric and fusion coefficients:
$$\langle i | j \rangle = \delta_{ij}, \quad \langle i | j ,k \rangle = N_{ijk}.$$ 
Topologically, the first equation corresponds to gluing two solid tori, each containing a Wilson line in representation $i$ and $j$, respectively. As a result, one gets a 3-manifold $S^{1} \times S^{2}$ containing two Wilson lines in $i,j$ wrapping the $S^{1}$ direction. LHS stands for the partition function of this configuration. Now, this partition function is nothing but the trace on the associated Hilbert space on $S^{2}$ with two punctures decorated by $i$ and $j$. The charge conservation argument immediately tells us that $i$ and $j$ must be dual to each other. For us, $i$ and $j$ are spin representations, and thus it is enough to write this condition as $\delta_{ij}$. The second equation corresponds to gluing two solid tori, and now one solid torus contains one Wilson line in $i$ representation, while the other contains two Wilson lines in $j$ and $k$. Again, the resultant 3-manifold is $S^{1} \times S^{2}$, and the partition function is nothing but the trace on the Hilbert space associated to $S^{2}$ with three punctures decorated by $i,j,k$. Again from the charge conservation argument, we see that the RHS must be the fusion coefficient among the spin representations $i,j$ and $k$.

Next, let us briefly recall the central ideas of \cite{BFLP}. Consider Wilson loop operators supported on a $m$-componnet link $\mathcal{L} = \bigcup_{i=1}^{m} C_{i}$. Each component $C_{i}$ is colored by an irreducible representation $R_{i}$ of $SU(2)$, and the link lies in $S^{3}$. The complement of $\mathcal{L}$, $\mathcal{L}^{c} = S^{3}\setminus D \times \mathcal{L}$ is obtained by removing a small solid torus $D \times \mathcal{L}$ in $S^{3}$. Since $\mathcal{L}^{c}$ is a 3-manifold whose boundary is a disjoint union of $m$ tori, a path integral on $\mathcal{L}^{c}$ fixes a vector in $(H_{T^{2}})^{\otimes m}$:

$$|\mathcal{L} \rangle = \sum_{\alpha_{1}, \cdots, \alpha_{m}} C(\alpha_{1},\cdots,\alpha_{m}) | \alpha_{1} \rangle \otimes \cdots \otimes |\alpha_{m} \rangle,$$
where $\alpha_{i}$'s are the spin representations which span $H_{T^{2}}$. Taking an inner product with a fixed vector $\langle \alpha_{1} | \otimes \cdots \otimes \langle \alpha_{m} |$, we are effectively gluing $m$ solid tori back into $S^{3}$, but this time each solid torus $D \times C_{i}$ contains a Wilson line colored in $\alpha_{i}$. As a result, on LHS we get the expectation value $\langle W_{\alpha_{1}, \cdots, \alpha_{m}}(\mathcal{L}) \rangle$. Since the metric on $H_{T^{2}}$ is nothing but a Kronecker delta symbol, we get on the RHS the coefficient $C(\alpha_{1},\cdots,\alpha_{m})$. Thus, we see that the link state is nothing but the sum over the basis vectors with the expectation value of Wilson loops as the coefficients:

$$|\mathcal{L} \rangle = \sum_{\alpha_{1}, \cdots, \alpha_{m}} \langle W_{\alpha_{1},\cdots,\alpha_{m}}(\mathcal{L})\rangle |\alpha_{1} \rangle \otimes \cdots \otimes |\alpha_{m} \rangle.$$

Once the colored link invariants are known, we can explicitly write down the $m$-partite entangled state corresponding to $\mathcal{L}$. Then, the entanglement structure of $|\mathcal{L}\rangle$ can be studied by computing its (reduced) density matrix, entanglement entropy, or entanglement negativity. In \cite{BFLP}, several examples were provided, including the triple Hopf link $2^{2}_{1} + 2^{2}_{1}$ and the Borromean ring $6^{3}_{2}$ (both in Rolfsen notation.) The corresponding link states can be explicitly written in terms of modular $S$ and $T$ matrices as follows, in which case the expectation values are colored Jones polynomials:

\begin{gather*}
| 2^{2}_{1} + 2^{2}_{1} \rangle = \sum_{j_{1},j_{2},j_{3}} \frac{S_{j_{1}j_{2}}S_{j_{2}j_{3}}}{S_{0j_{2}}} | j_{1}\rangle \otimes |j_{2} \rangle, \quad S_{j_{1}j_{2}} = \sqrt{\frac{2}{k+2}}\sin \big( \frac{(2j_{1}+1)(2j_{2}+1) \pi}{k+2} \big), \\
| 6^{3}_{2} \rangle = \sum_{i=0}^{\min(j_{1},j_{2},j_{3})} (-1)^{i}(q^{1/2}-q^{-1/2})^{4i} \frac{[2j_{1}+i+1]![2j_{2}+i+1]![2j_{3}+i+1]![i]![i]!}{[2j_{1}-i]![2j_{2}-i]![2j_{3}-i]![2i+1]![2i+1]!}, \\
\text{where} \quad q = e^{\frac{2 \pi i}{k+2}}, \quad [n] = \frac{q^{n/2}-q^{-n/2}}{q^{1/2}-q^{-1/2}}, \quad \text{and} \quad [n]! = [n][n-1] \cdots [1].
\end{gather*}

The density matrix for $| \mathcal{L} \rangle$ is, as usual, $\rho_{\mathcal{L}} = \frac{1}{\langle \mathcal{L} | \mathcal{L} \rangle} |\mathcal{L} \rangle \langle \mathcal{L} |$. Tracing over components of $\mathcal{L}$, we get the reduced density matrices. In the above two examples, tracing out any one component of the triple Hopf link yields a separable reduced density matrix, indicating that the state $|2^{2}_{1}+2^{2}_{1} \rangle$ is ``GHZ-like''. On the other hand, tracing out any one component of the Borromean ring yields a non-separable density matrix, showing that the state $| 6^{3}_{2} \rangle$ is ``W-like''.

\section{Review: Colored link invariants via symmetric webs}
The triple Hopf link is a torus link, so its colored link invariants can be written in terms of modular $S$ and $T$ matrices of $\widehat{su(2)}_{k}$ WZW model. The Borromean ring is a twist ink, and Habiro's formula works nicely to compute its colored link invariants. To compute colored link invariants for more general classes of knots/links, one may introduce junctions of Wilson lines and apply the techniques of quantum spin networks \cite{Masbaum, MasbaumVogel, CGV}. Alternatively, we may resolve the crossings by the ``symmetric webs'', which we will soon discuss.

In subsection \ref{subsec:local}, we review networks of Wilson lines and the local relations among them \cite{Witten89, Witten89wf, Witten89rw, MOY, CKM, CGR}. In the following subsection, we discuss their symmetric analogues \cite{RoseTub, ChunRefinedCS}. We provide an examples of the figure-eight knot and the triple Hopf link when $k=2$. The example will allow us to explicitly write down the entangled link state.

\subsection{Local relations among Wilson lines and networks of Wilson lines}
\label{subsec:local}
As was discussed before, path integral over a 3-manifold $M_{3}$ with boundary fixes a vector in the associated Hilbert space, $H_{\partial M_{3}}$. The Hilbert space is isomorphic to the space of conformal blocks in $\widehat{su(2)}_{k}$ WZW model on $\partial M_{3}$. When $M_{3}$ contains Wilson lines which end on the boundary, the Hilbert space is isomorphic to the space of conformal blocks on $\partial M_{3}$ with punctures decorated by the $R_{1}, \cdots, R_{m}$, the representations Wilson lines carry. 

In particular, consider the case when $M_{3} = D^{3}$, \textit{i.e.}, the closed 3-ball. The boundary $\partial M_{3}$ is simply a 2-sphere, and thus, the charge conservation argument shows that the dimension of the corresponding Hilbert space is equal to the dimension of the invariant subspace :

\begin{equation}
dim \, H_{S^{2} ; R_{1}, \cdots, R_{m}} = dim \, \mathrm{Inv}_{G}(\otimes_{i=1}^{m} R_{i}).
\label{eqn:cc}
\end{equation}
Now, let $dim \, H_{S^{2} ; R_{1}, \cdots, R_{m}} = d$, and consider $(d+1)$ distinct Wilson line configurations in $D^{3}$ ending on the boundary 2-sphere with $m$ punctures $R_{1}, \cdots, R_{m}$. Then, $(d+1)$ vectors obtained by the path integral satisfy a linear relation in the $d$-dimensional Hilbert space. 

One canonical example is the famous skein relation, Figure \ref{fig:skein}. In Figure \ref{fig:skein}, the associated Hilbert space $H_{S^{2} ; \square, \square, \bar{\square}, \bar{\square}}$ is 2-dimensional (by Equation \ref{eqn:cc}). The three distinct braided/unbraided Wilson lines in $D^{3}$ then fix three vectors in the two-dimensional Hilbert space, and they clearly satisfy a linear relation. The coefficients of the linear relation is determined from the eigenvalues of the modular $T$ matrix, as was explained in \cite{Witten89}. Using the skein relation, we can simplify any given $\square$-colored link until there is no crossing left and write its expectation value in terms of those of unknots.

For links colored in higher spin representations, however, the Hilbert space $H_{S^{2} ; j_{1}, j_{2}, \bar{j_{1}}, \bar{j_{2}}}$ is $\min (j_{1},j_{2})+1$ dimensional. We would need $\min (j_{1},j_{2})+2$ braided Wilson lines to set up a linear relation, but such a linear relation may not simplify the given knot, as we are adding more crossings. Thus, we need an alternative way to simplify the given link into a form that we can evaluate systematically.

\subsection{Symmetric webs in $SU(2)$ Chern-Simons theory}
\label{subsec:symweb}

We can do so by introducing jucntions of Wilson lines and resolve the crossings with trivalent graphs of Wilson lines \cite{Witten89wf, Witten89rw}. The junctions of interest are trivalent, and on each of them we place a gauge invariant tensor so that a closed trivalent graph defines a gauge invariant observable.

\begin{figure} [htb]
\centering
\includegraphics{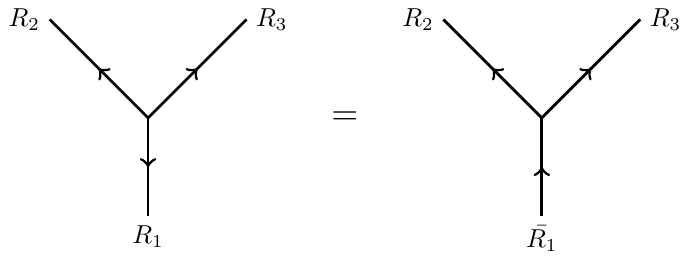}
\caption{LHS: a junctions of three Wilson lines colored in $R_{1}, R_{2}, R_{3}$ such that $0 \in R_{1} \otimes R_{2} \otimes R_{3}$. At the junction, we place a gauge invariant tensor in $Hom_{G}(R_{1} \otimes R_{2} \otimes R_{3}, \mathbb{C})$. RHS: an equivalent junction, with $R_{1}$-strand reversed and replaced by its complex dual.}
\label{fig:MOYjunction}
\end{figure}

When Wilson lines are colored by antisymmetric powers of the fundamental representations of $SU(N)$, such trivalent graphs coincide with ``MOY graphs'' \cite{CGR}. The MOY graphs can be simplified systematically by local relations, until they can be written as a linear sum of MOY graphs whose ``MOY graph polynomials'' are known \cite{MOY}. The networks of Wilson lines in antisymmetric representations satisfy the same set of local relations, which are also called $N$Web relations in the context of representation theory of quantum groups \cite{CKM}. 

Before proceeding further, it is important to note that the Wilson lines with junctions must be vertically framed. This is because we cannot canonically frame the Wilson lines near the junctions so that the configuration's self-linking number to vanish upon braiding. Although the colored link invariants of vertically framed Wilson lines are different from those which are canonically framed, the entanglement structure would be framing-independent \cite{BFLP}. For this reason, we fix the framing of Wilson lines to be vertical in the rest of this paper.

Now let us consider the Wilson lines colored in spin representations. For $SU(2)$, the spin-$i/2$ representation is simply the $i$-th symmetric power of fundamental representations, denoted $Sym^{i}\square$. Just like their antisymmetri counterparts, they constitute ``symmetric webs'' which enable us to compute the colored link invariants by replacing the crossings with planar trivalent graphs of Wilson lines. The key trick is to use the level-rank duality in 2d WZW models, in which we can swap $\wedge^{i}\square$ with $Sym^{i}\square$ and the rank of the gauge group $N$ with the level $k$. To see how this works, consider the expectation value of a Wilson loop colored in $Sym^{i}\square$:

$$\langle W_{S^{i}\square}(\text{unknot}) \rangle = S^{(N,k)}_{0 Sym^{i}\square}/S^{(N,k)}_{00} = \quantumbinomial{N+i-1}{i},$$
where the superscript $(N,k)$ indicates that the S-matrix is from $\widehat{su(N)}_{k}$ WZW model. One can in fact write the RHS in terms of the S-matrix elements from the $\widehat{su(k)}_{N}$ WZW model:

\begin{gather}
q^{N} = e^{\pi i N/(N+k)} = -e^{\pi i k / (N+k)} = -q^{k} \\[1.5ex]
\Rightarrow \quad [N+a] = \frac{q^{N+a}-q^{-(N+a)}}{q-q^{-1}} = \frac{q^{k-a}-q^{-(k-a)}}{q-q^{-1}} = [k-a] \\[1.5ex]
\Rightarrow \quad \quantumbinomial{N+i-1}{i} = \quantumbinomial{k}{i} = S^{(k,N)}_{0 \wedge^{i}\square}/S^{(k,N)}_{00},
\end{gather}
where in the last line we have introduced quantum binomials, $\quantumbinomial{a}{b} = \frac{[a][a-1]\cdots[a-b+1]}{[b][b-1]\cdots[1]}$.

\begin{figure} [htb]
\centering
(circle removal) \quad \raisebox{-0.5\height}{\includegraphics{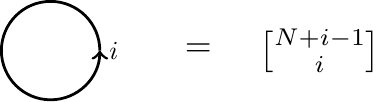}} \\[1.5ex]
(digon removal) \quad \raisebox{-0.5\height}{\includegraphics{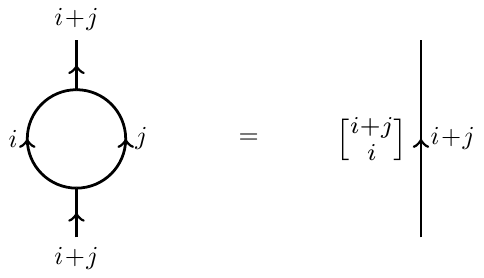}} \\[1.5ex]
(associativity) \quad \raisebox{-0.5\height}{\includegraphics{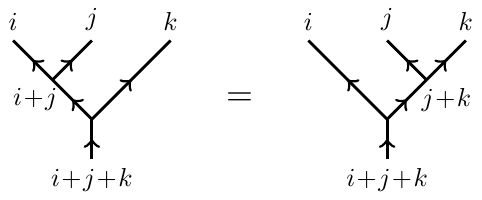}} \\[1.5ex]
($[E,F]$ relation) \quad \raisebox{-0.5\height}{\includegraphics{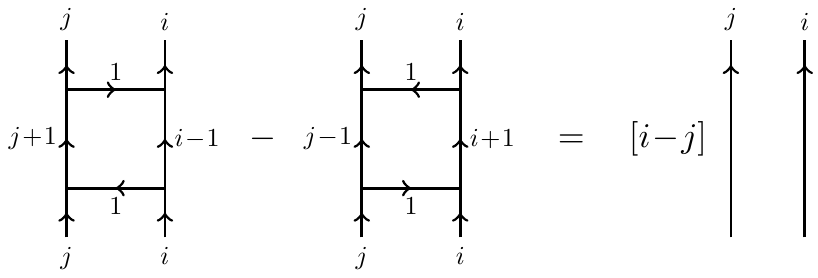}}
\caption{The local relations of Wilson lines in symmetric representations, which determine the expectation values of all closed trivalent graphs colored in symmetric representations. Above, the indices $i,j,k$ stand for $Sym^{i}\square, Sym^{j}\square, Sym^{k}\square$, respectively.}
\label{fig:SymWebGenerators}
\end{figure}

From the dimension of the Hilbert space associated to $S^{2}$ with punctures, it is immediate that the kinematics of the Wilson lines in symmetric representations are the same as those of their antisymmetric counterparts. Therefore, we can obtain the symmetric web relations starting from the $k$Web relations and replacing $q^{k}$ by $-q^{N}$ and antisymetric representatoins with symmetric representations (Figure \ref{fig:SymWebGenerators}).

\begin{figure} [htb]
\centering
\includegraphics{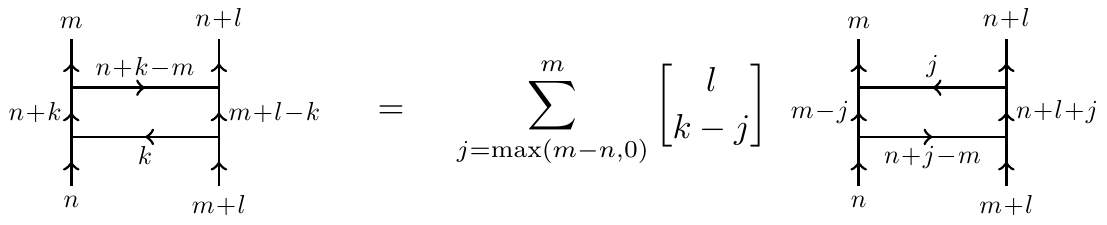}
\caption{The ``square switch'' relation.}
\label{fig:ss}
\end{figure}

The symmetric web relations provided in Figure \ref{fig:SymWebGenerators} are coherent and allow us to determine the expectation value of all closed trivalent graphs of Wilson lines colored in symmetric representations. Applying $[E,F]$ relation repeatedly, we can derive the famous ``square switch'' relation, which is particularly useful when simplifying complicated Wilson line networks (Figure \ref{fig:ss}).

\begin{figure} [htb]
\centering
\includegraphics{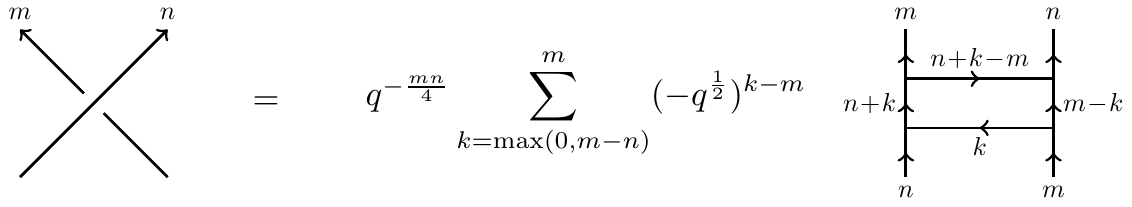}
\includegraphics{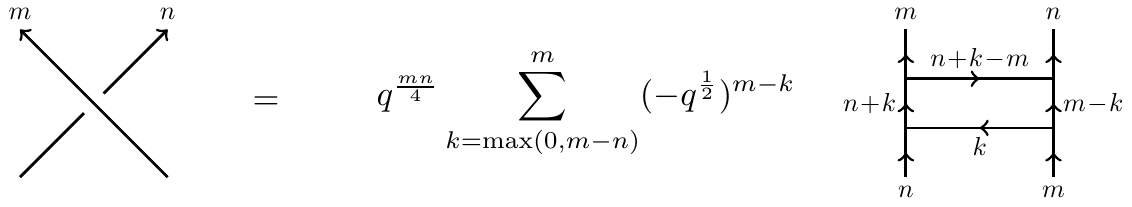}
\caption{Resolution of crossings.}
\label{fig:crossingResolutions}
\end{figure}
Now it remains to write the crossings of symmetric-colored Wilson lines as a linear sum of trivalent graphs. In Figure \ref{fig:crossingResolutions}, the Wilson lines in the LHS and RHS satisfy a linear relation because $H_{S^{2};Sym^{m}\square, Sym^{n}\square, \overline{Sym^{m}\square}, \overline{Sym^{m}\square}}$ is $(\min(m,n)+1)$-dimensional. It is rather tedious to fix the coefficients, so we refer the interested readers to the appendix \ref{appendix:crossing} for the derivation of the above relations.

So far, we have restricted ourselves to ``MOY'' type junctions. Alternatively, one may generalize to include more general types of junctions, those which appear in the quantum spin networks. When incoming three Wilson lines are colored in spin $i$, $j$ and $k$, we insert a vertex if and only if the fusion coefficient $N_{ijk}$ is nonzero. These are precisely the types of junctions considered in \cite{Witten89wf, Witten89rw}, and one obtains quantum spin networks \cite{Masbaum,MasbaumVogel,CGV} by choosing a different normalization for the gauge invariant tensors from those of \cite{Witten89wf, Witten89rw}.

\subsection{Example: Triple Hopf link}
Now let us provide some examples in which the link states are computed via the symmetric webs. The first example is the simplest of all four-component link, the triple Hopf link.

For simplicity of calculation, we restrict ourselves to $k=2$. This is because when $k=1$ the Wilson lines can only be colored by trivial and fundamental representations, and then we can compute the link invariants simply via skein relations. For higher level $k$, the complexity will grow with it, but the idea is the same: resolve all crossings and use the symmetric web relations to simplify the resultant trivalent graphs. 

\begin{figure} [htb]
\centering
\includegraphics{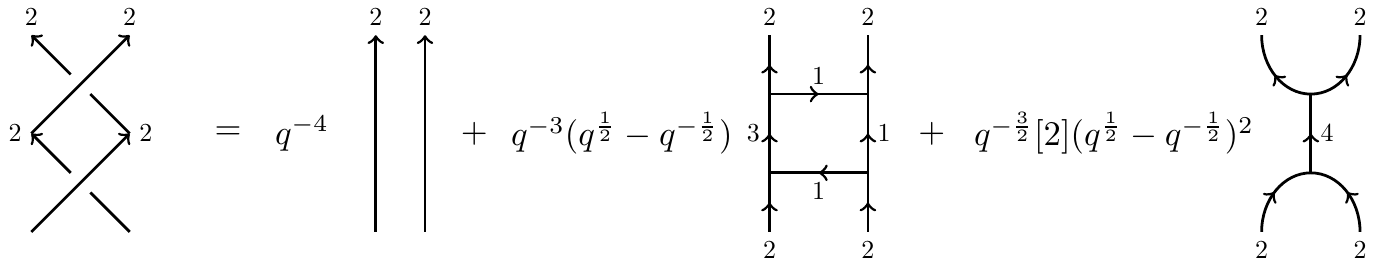}\\[1.5ex]
\includegraphics{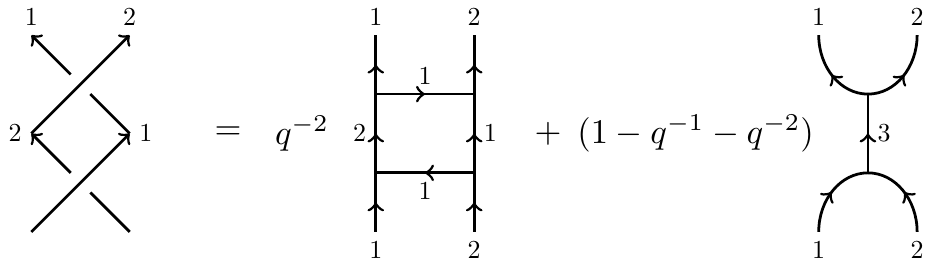}\\[1.5ex]
\includegraphics{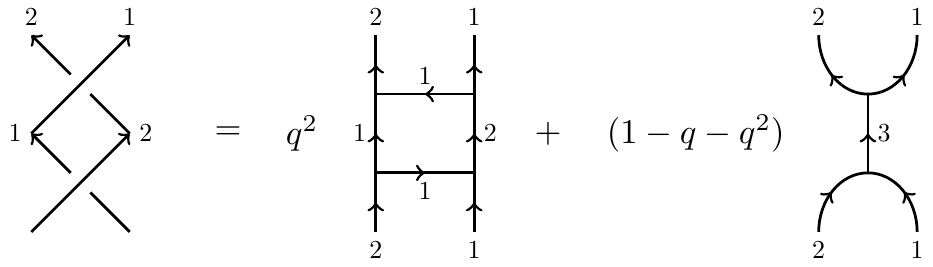}\\[1.5ex]
\includegraphics{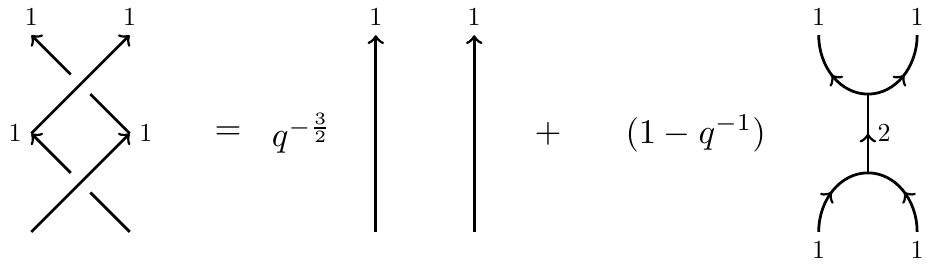}
\caption{The relations among Wilson lines which are used to the ``linkings'' of $1$ and $2$-colored triple Hopf links.}
\label{fig:doubleCrossings}
\end{figure}

\begin{figure} [htb]
\centering
\includegraphics{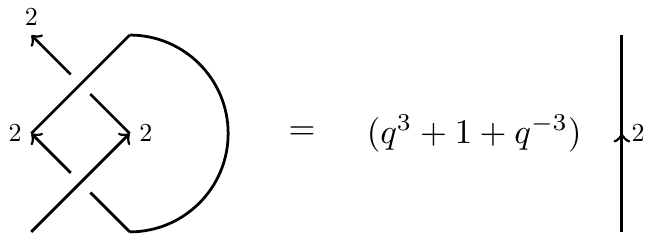} \quad \quad
\includegraphics{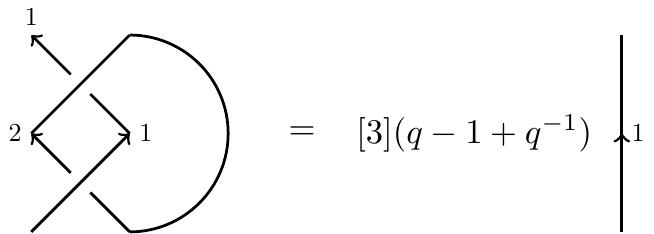} \\[1.5ex]
\includegraphics{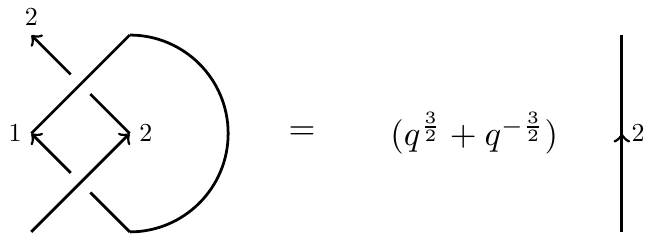}\quad \quad
\includegraphics{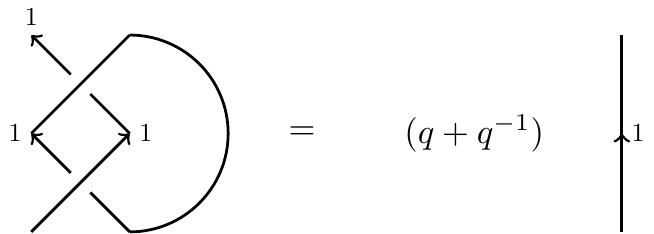}
\caption{The relations obtained by closing one of the Wilson lines from the previous figure.}
\label{fig:doubleCrossingsClosed}
\end{figure}

\begin{figure} [htb]
\centering
\includegraphics{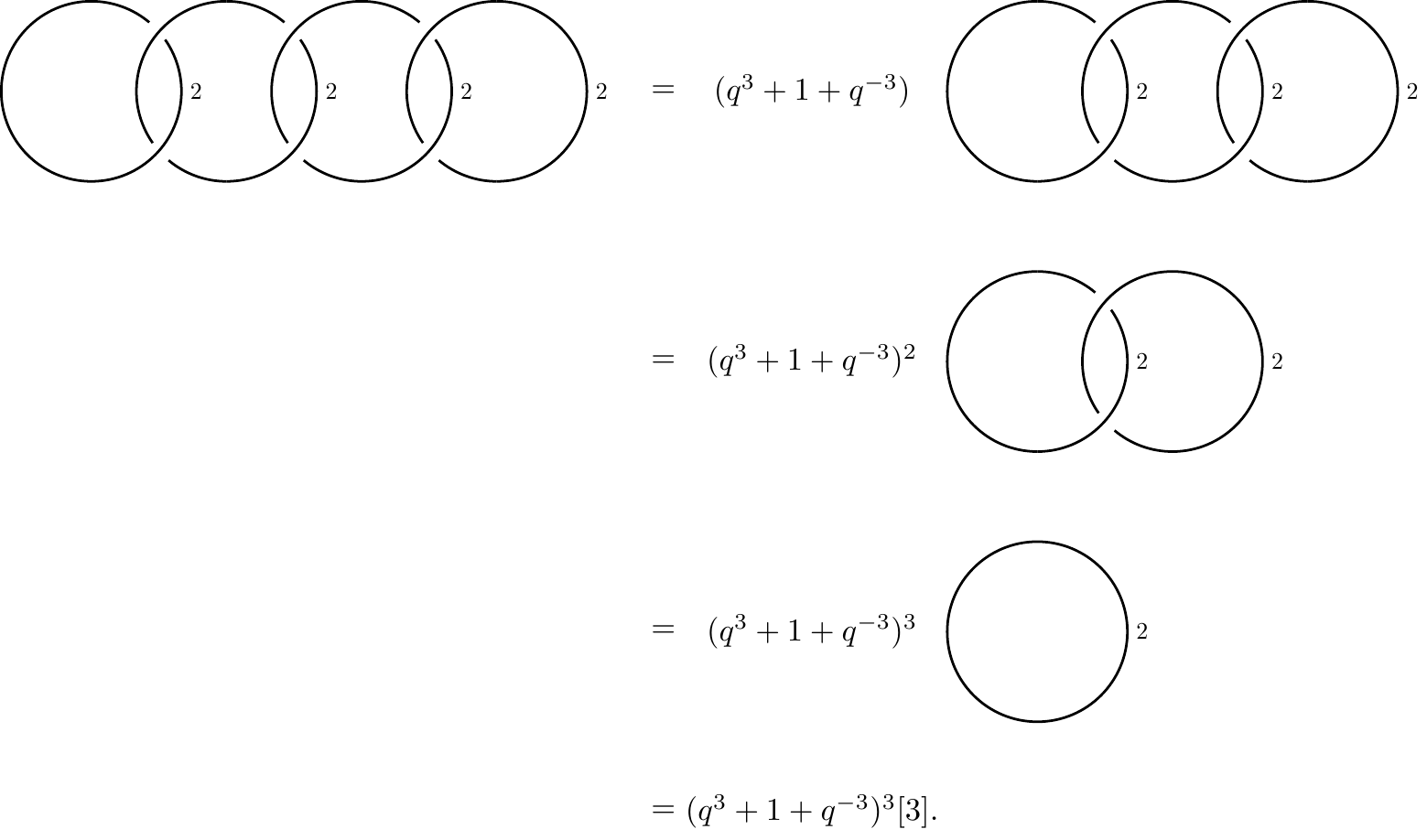}
\caption{Evaluation of $J_{2,2,2}(2^{2}_{1}+2^{2}_{1}+2^{2}_{1})$ via symmetric web relations.}
\label{fig:eval222tripleHopf}
\end{figure}

In Figure \ref{fig:doubleCrossings}, the Wilson lines on the LHS are obtained by vertically stacking the resolution of crossings, Figure \ref{fig:crossingResolutions}. Closing a Wilson line, we get the relations in Figure \ref{fig:doubleCrossingsClosed}. Given a colored triple Hopf link, we can apply them from right to left, until we are left with an unknot. For instance, consider a triple Hopf link, whose components are all colored by $Sym^{2}\square$. Applying the topmost relation of Figure \ref{fig:doubleCrossingsClosed} from right to left twice, we can compute the colored link invariant (see Figure \ref{fig:eval222tripleHopf}). Notice that we have omitted the orientation of Wilson lines, as the spin reprsentations are self-dual.

Likewise, we can compute the colored link invariants of the triple Hopf link for all colorings up to $Sym^{2}\square$ and determine the link state of the triple Hopf link at level $k=2$. Here, we use the fact that $q = e^{2 \pi i /(2+2)} = i$ to simplify this 64-dimensional vector. The full $q$-dependent expression is provided in Appendix \ref{appen:qTripleHopfState}.

\begin{align*}
| 2^{2}_{1}+2^{2}_{1}+2^{2}_{1} \rangle \quad = \quad &|0000\rangle + \sqrt{2} \big( |1000\rangle + |0100\rangle + |0010\rangle + |0001\rangle \big) \\
&+ \big( |2000\rangle + |0200\rangle + |0020\rangle + |0002\rangle \big) \\
&+ 2 \big( |1010\rangle + |0101\rangle + |1001\rangle \big) \\
&+ \sqrt{2} \big( |2010\rangle + |1020\rangle + |0201\rangle + |0102\rangle + |2001\rangle + |1002\rangle \big) \\
&+ \big( |2200\rangle + |0220\rangle + |0022\rangle + |2020\rangle + |2002\rangle + |0202\rangle \big) \\
&+ (-2) \big( |2101\rangle + |1201\rangle + |1021\rangle + |1012\rangle \big) \\
&+ \sqrt{2} \big( |2201\rangle + |1022\rangle \big) -\sqrt{2} \big( |1202\rangle + |2102\rangle + |2021\rangle + |2012\rangle \big) \\
&+ \big(|2202\rangle + |2022\rangle \big) + 2 \big( |0121\rangle + |1210\rangle \big) \\
&+  \sqrt{2} \big( |0212\rangle + |2120\rangle \big) -\sqrt{2} \big( |0122\rangle + |1220 \rangle + |0221\rangle + |2210 \rangle \big) \\
&+ \big( |0222\rangle + |2220\rangle \big) -\sqrt{2} \big( |2121\rangle + |1212\rangle \big) + 2 |1221\rangle \\
&+ \sqrt{2} \big( |2122\rangle + |2212\rangle \big) -\sqrt{2} \big( |1222\rangle + |2221\rangle \big) + |2222\rangle.
\end{align*}
where each cubit corresponds to a component of the triple Hopf link, and the number assigned to the cubit is its coloring.

\subsection{Example: Figure-eight knot}
Next, we consider our first example of a non-torus knot/link. The simplest of non-torus knot/link is the figure-eight knot: the simplest in a sense that it has the minimum number of crossings. Since there is only one component, the link state is 3-dimensional, and the coefficients of $|0\rangle$ and $|1\rangle$ can be easily computed by applying the skein relation in vertical framing. Let us omit the detailed procedure and write them down as follows:

$$ 1 | 0 \rangle, \quad \text{and} \quad (q^{2}+1-q)(q^{\frac{1}{2}}+q^{-\frac{1}{2}}) | 1\rangle.$$

\begin{figure} [htb]
\centering
\includegraphics{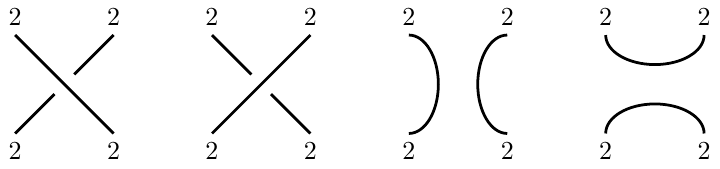}
\caption{Four 2-colored Wilson lines which satisfy a linear relation inside a 3-ball.}
\label{fig:22_lines}
\end{figure}

Now, it remains to compute the colored link invariant of a $Sym^{2}\square$-colored figure-eight knot. We can compute the link invariant most easily by using a linear relation among four 2-colored Wilson lines shown in Figure \ref{fig:22_lines} (such a linear relation is well-defined, for $H_{S^{2},2,2,\bar{2},\bar{2}}$ is 3-dimensional.)

\begin{figure} [htb]
\centering
\includegraphics{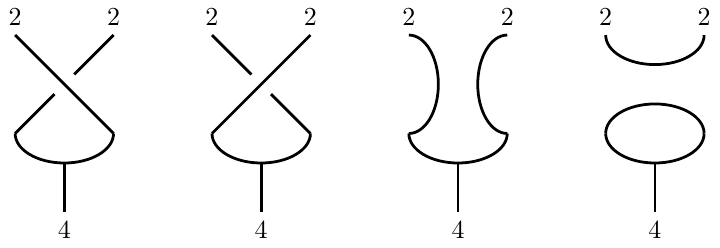} \\[1.5ex]
\includegraphics{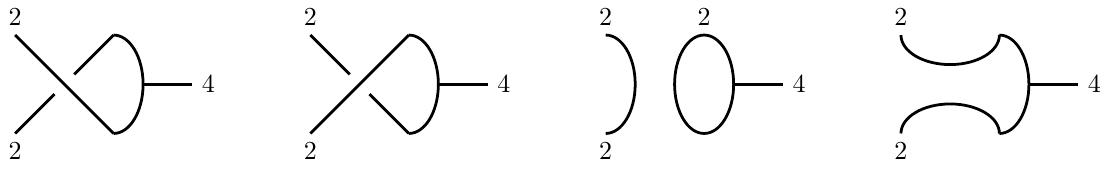}\\[1.5ex]
\includegraphics{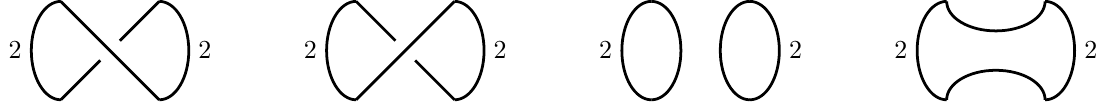}\\[1.5ex]
\includegraphics{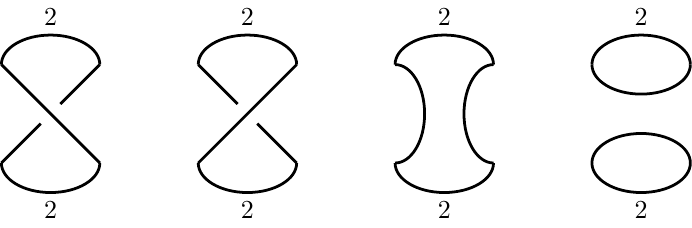}
\caption{Four different ways to ``close'' the 2-colored Wilson lines satisfying a linear relation.}
\label{fig:22_more_lines}
\end{figure}

And symmetric web relations allow us to determine coefficients of the linear relation of our interest, by considering the four Wilson lines in Figure \ref{fig:22_lines} as parts of the Wilson lines shown in Figure \ref{fig:22_more_lines}. 

The ``tadpole'' diagrams in the first two closures vanish as they violate the charge conservation condition. Then, the braid relations in Figure \ref{fig:braid} allow us to determine the coefficients as shown in Figure \ref{fig:22_lines_coeff}. Apply the linear relation to any one crossing in a figure-eight knot. Then, we can write its colored link invariant as a linear sum over an unknot, a trefoil knot and a Hopf link. The $|2\rangle$ component is therefore:

\begin{figure} [htb]
\centering
\includegraphics{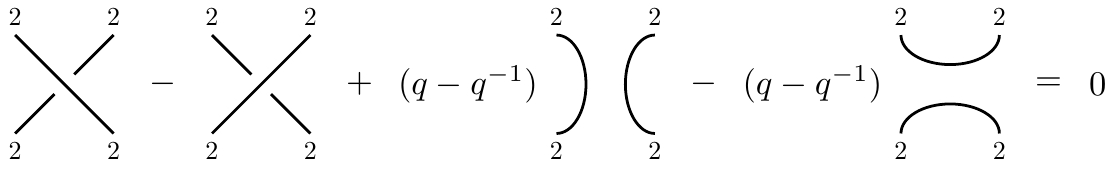}
\caption{A linear relation among four 2-colored Wilson lines.}
\label{fig:22_lines_coeff}
\end{figure}

$$ (q^{7}-q^{5}+q+1+\frac{1}{q}-\frac{1}{q^{5}} +\frac{1}{q^{7}}) | 2 \rangle.$$
When $k$ equals 2, $q = e^{2 \pi i /(2+2)}$, and the link state of a figure-eight knot can be explicitly written:

$$|4_1 \rangle = |0 \rangle  - \sqrt{2}i |1 \rangle + |2\rangle.$$

\section{Conjecture: entanglement structure and topological entanglement}
\label{sec:entangled}

Consider a link which is a union of two unlinked sub-links. TQFT axioms state that the corresponding link state must be a product state. But is the converse true? That is, given a link state which is a product state, can we expect the link itself to be a union of unlinked components? 

If a link state $|\mathcal{L} \rangle$ is a product of two component link states, $|\mathcal{L} \rangle = |\mathcal{L}_{1} \rangle \otimes |\mathcal{L}_{2} \rangle$, any operator that acts on either $|\mathcal{L}_{1}\rangle$ or $|\mathcal{L}_{2}\rangle$ (as $1 \otimes \hat{O}$ or $\hat{O} \otimes 1$) would not change the other. In particular, a surgery on one component (say, $\mathcal{L}_{1}$) would not change the state $|\mathcal{L}_{2}\rangle$. The topological implication of the claim $|\mathcal{L}\rangle = |\mathcal{L}_{1}\rangle \otimes |\mathcal{L}_{2}\rangle$ can be best illustrated when the gauge group is $U(1)$. In this case, the (colored) link invariants are nothing but (colored) Gaussian linking nubmers, and the above claim implies that the ``mutual linking number'' (analogous to the mutual inductance) between $\mathcal{L}_{1}$ and $\mathcal{L}_{2}$ vanishes and remains unaltered after arbitrary surgery on either of the component links. In terms of the (colored) link invariants, this implies that the colored link invariants of $\mathcal{L}$ factorize into those of $\mathcal{L}_{1}$ and $\mathcal{L}_{2}$.

Obviously, the claim would be true if $\mathcal{L}_{1}$ and $\mathcal{L}_{2}$ are unlinked. However, the linking number is not an exclusive linking detector (for instance, the whitehead link has a vanishing linking number, but its component are nontrivially linked), so the condition does not immediately enforce $\mathcal{L}$ to be a product link. Currently, we do not have a proof or counterexample, so we leave it as a conjecture here. When the gauge group is $SU(2)$:

\begin{conj}
Given a $m$-component link $\mathcal{L}$, suppose there exist two sub-links $\mathcal{L}_{1}$ and $\mathcal{L}_{2}$, each with $i$ and $(m-i)$ components. Suppose the two sub-links satisfy the following:
$$J_{\alpha_{1}, \cdots, \alpha_{m}}(\mathcal{L}) = J_{\alpha_{1}, \cdots, \alpha_{i}}(\mathcal{L}_{1})J_{\alpha_{i+1}, \cdots, \alpha_{m}}(\mathcal{L}_{2})$$ 
for all colorings $\alpha_{1}, \cdots, \alpha_{m}$, then $\mathcal{L}_{1}$ and $\mathcal{L}_{2}$ are unlinked.
\end{conj} 

An information theoretic plausiblity argument for the above would go as follows. If the link state  $|\mathcal{L} \rangle$ is of form  $|\mathcal{L} \rangle = |\mathcal{L}_{1} \rangle \otimes |\mathcal{L}_{2} \rangle$ then a partial trace over either the  $|\mathcal{L}_{1} \rangle$ or $|\mathcal{L}_{2} \rangle$ subsystem would not affect the other. In particular, tracing out one would leave the other invariant. At this point, one can simply construct the link corresponding to $|\mathcal{L}_{1} \rangle$ and $|\mathcal{L}_{2} \rangle$ separately, and simply put them next to each other at the end of the construction to create an unlinked manifestation of $|\mathcal{L} \rangle$. This of course gives plausibility to the fact that they are genuinely not linked.

\section{Categorification of the entanglement entropy and the density matrix}
A ``categorification'' is an algebraic procedure in which algebraic objects are upgraded to the higher level ones, possibly with further structures: \textit{e.g.}, numbers to vector spaces, vector spaces to categories, and $n$-categories to higher $(n+1)$-categories. When topological invariants admit categorification, they often produce strictly stronger invariants. For instance, $\mathfrak{sl}_{N}$ polynomials are categorified to Khovanov or Khovanov-Rozansky homologies \cite{Kh, KhR04, KhR05}, which can distinguish knots and links better than the former.

Now that the symmetric webs are originally defined as a 1-category \cite{RoseTub}, a map (\textit{e.g.}, a density matrix) between symmetric webs would necessarily be a 2-morphism in the categorification of symmetric webs. In fact, we have seen a 2-morphism between symmetric webs, the density matrix! Recall that when computing the colored link invariants, we have replaced each crossings by certain symmetric webs which belong to the same Hilbert space. One can quickly notice that the resolution of crossing is nothing but a basis change inside the Hilbert space. Then, the density matrix can be interpreted as a map between two symmetric webs, or in the language of higher representation theory, a 2-morphism between symmetric webs. 

Unfortunately, the diagrammatic presentation of such 2-category (called the symmetric $\mathfrak{sl}_{2}$-foam 2-category in \cite{RoseTub}) is unknown at present, but we can still take a glimpse of what it would look like from its antisymmetric counterpart. The networks of Wilson lines colored in antisymmetric representations correspond to morphisms in the category of $\mathfrak{sl}_N$-webs \cite{CKM, CGR}. When categorified, the morphisms between webs are represented by singular cobordisms connecting them \cite{BN, Blanchet, CMW, Kh03, KhR04, Kup, MV, SN, LQR, QR}. These cobordisms satisfy certain relations, which encode homological information which is not fully captured in the webs.

Thus, once the symmetric $\mathfrak{sl}_{2}$-foam 2-category is successfully constructed, the density matrix would correspond to a linear sum of singular cobordisms between symmetric webs. Then, the kinematics of the symmetric $\mathfrak{sl}_{2}$-foams would allow us to study the entanglement structure of the link states not only at the level of knot polynomials, but also in terms of the homological invariants.

\section{Conclusions}
In this paper we have married the techniques developed in \cite{BFLP,Salton:2016qpp} with symmetric web techniques to create a novel way of creating topologically interesting states. Further, we make and motivate a conjecture that product states are represented by un-linked knots.

It would be interesting to attempt to prove this conjecture in future work. Also, now that this connection has been established between entanglement and knot aspects of topological theories  it would be intresting if some other more refined entanglement property could be useful in defining the cohomologies in the categorification of topological quantum field theories.

\acknowledgments{
We would like to thank Fran\c{c}ois Costantino, Sergei Gukov, Peter Kravchuck, Greg Kuperberg, and Onkar Parrikar for discussions. N.B. is funded as a Burke Fellow at the Walter Burke Institute for Theoretical Physics. The work is funded in part by the DOE Grant DE-SC0011632 and the Walter Burke Institute for Theoretical Physics, and also by the Samsung Scholarship.}

\appendix
\section{Derivation of the crossing resolution formula}
\label{appendix:crossing}
\begin{figure} [htb]
\centering
\includegraphics{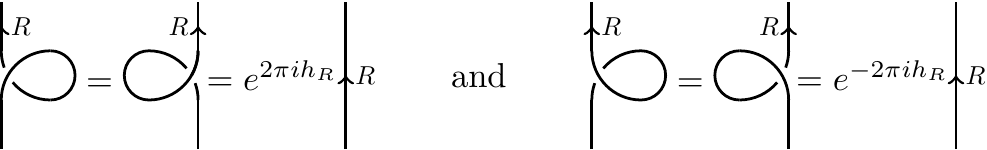}
\includegraphics{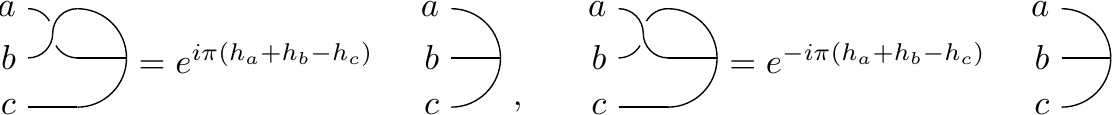}\\[1.5ex]
\includegraphics{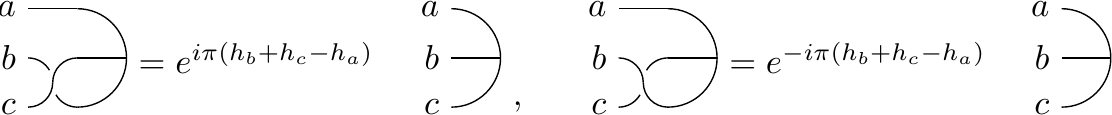}
\caption{Braiding relations of vertically framed Wilson lines.}
\label{fig:braid}
\end{figure}

In this appendix, we derive the crossing resolution formula, Figure \ref{fig:crossingResolutions}. First of all, since our Wilson lines are vertically braided, they are subject to braid relations in Figure \ref{fig:braid}. In Figure \ref{fig:braid}, Wilson lines are colored by the integrable representations of $\widehat{su(2)}_{k}$, $a,b,c$ and $R$, and $h_{a},h_{b},h_{c},h_{R}$ are the conformal weights of the corresponding primary fields in $\widehat{su(2)}_{k}$ WZW model on a punctured $2$-sphere. Explicitly, they are given by:

$$e^{\pi i h_{R}} = e^{\frac{\pi i }{N+k} C_{2}(R)} = q^{\frac{C_{2}(R)}{2}}$$
where $C_{2}(R) = \frac{1}{2}(\kappa(R) + N |R| - \frac{|R|^{2}}{N})$ is the quadratic Casimir of the representation $R$ of $SU(N)$. Here, $|R|$ stands for the number of boxes in the Young tableau of $R$, and $\kappa(R) = |R| + \sum_{i} (R_{i}-2i)$, with $R_{i}$ representing the number of boxes in the $i$-th row of the Young tableau. For spin representations of $SU(2)$, we have $C_{2}(Sym^{i}\square) = \frac{i(i+2)}{4}$.

With the braid relations of Figure \ref{fig:braid} and the symmetric web relations Figure \ref{fig:SymWebGenerators}, we can derive the resolution of crossings formula by induction. Let us set up the base cases first.

\begin{figure} [htb]
\centering
\includegraphics{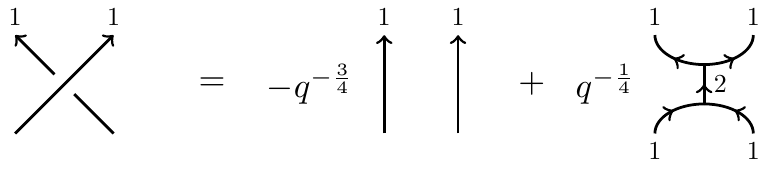} \\[1.5ex]
\includegraphics{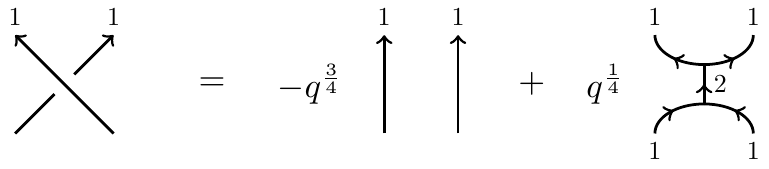}
\caption{Local relations which express braided Wilson lines in terms of planar graphs.}
\label{fig:base11}
\end{figure}

That the local relations must hold is evident from the dimension counting. The coefficients can also be determined by considering two different ways to ``close'' the Wilson lines. For instance, consider the top linear relation in Figure \ref{fig:base11}. We may close the Wilson lines in two different ways, as shown in Figure \ref{fig:11over_closure}. Using the braid relations and symmetric web relations, we can check that the shown coefficients are indeed correct. 

\begin{figure} [htb]
\centering
\includegraphics{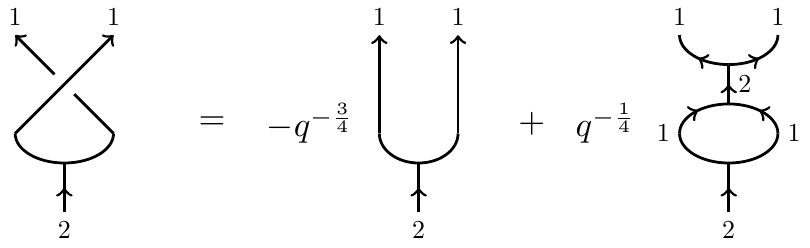} \\[1.5ex]
\includegraphics{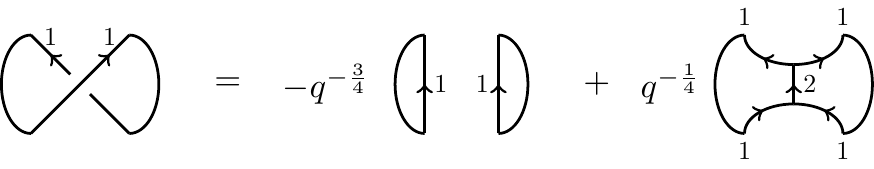}
\caption{Two different closures of the linear relation, in which crossed Wilson lines.}
\label{fig:11over_closure}
\end{figure}

Similarly, we may consider linear relations among the crossed Wilson lines with $m \geq 1$ and $1$ coloring and their planar resolutions. The coefficients of the linear relations are fixed in similar ways as in Figure \ref{fig:11over_closure}, by (1) inserting a junction below and (2) connecting the of the Wilson lines with itself. The resultant local relations are shown in Figure \ref{fig:base} and serve as the base cases for our induction.

\begin{figure} [htb]
\centering
\includegraphics{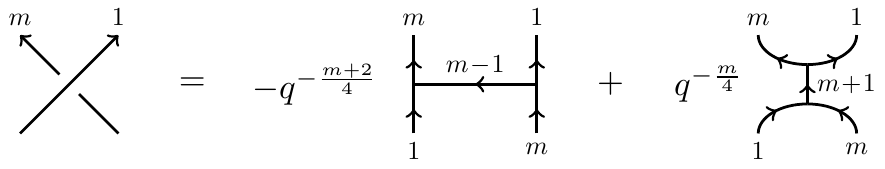} \\[1.5ex]
\includegraphics{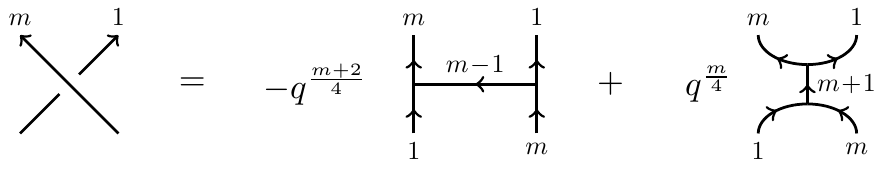}
\caption{Resolution of crossings when Wilson lines are colored by $m$ and $1$.}
\label{fig:11over_closure}
\end{figure}

\begin{figure} [htb]
\centering
\includegraphics{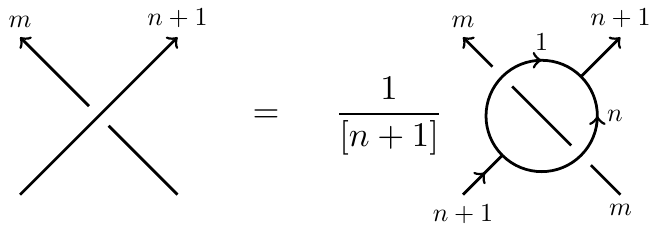} 
\caption{$(n+1)$-th induction step to derive the resolution of crossing formula.}
\label{fig:mn+1over_pop}
\end{figure}

Set up the induction hypothesis for crossings of $m$ and $n$-colored Wilson lines (as in Figure \ref{fig:crossingResolutions}) and consider crossings of $m$ and $(n+1)$-colored Wilson lines. Let us first consider the right-handed crossing. As in the case of $N$Webs and MOY graphs, we may first ``pop''  the $(n+1)$-colored Wilson line, as in Figure \ref{fig:mn+1over_pop}. Now we can apply the base case and the induction hypothesis on the resultant two crossings, and then the associativity and square switch relations of symmetric webs (Figure \ref{fig:SymWebGenerators} and Figure \ref{fig:ss}). The left-handed crossings can be derived likewise.

\section{The full $q$-dependent triple Hopf link state}
\label{appen:qTripleHopfState}

\begin{align*}
| 2^{2}_{1}+2^{2}_{1}+2^{2}_{1} \rangle \quad = \quad &|0000 \rangle + [2] \big( |1000 \rangle + |0100 \rangle + |0010 \rangle + |0001 \rangle \big) \\
&+ [3]\big( |2000 \rangle + |0200 \rangle + |0020 \rangle + |0002 \rangle \big) \\
&+ [2](q+q^{-1}) \big( |1100 \rangle + |0110 \rangle + |0011 \rangle \big) + [2]^{2}\big( |1010 \rangle + |0101 \rangle + |1001 \rangle \big) \\
&+ [2][3] \big( |2010 \rangle + |1020 \rangle + |0201 \rangle + |0102 \rangle + |2001 \rangle + |1002 \rangle \big) \\
&+ (q^{\frac{3}{2}}+q^{-\frac{3}{2}})[3] \big( |2100 \rangle + |0210 \rangle + |0021 \rangle + |1200 \rangle + |0120 \rangle + |0012 \rangle \big) \\
&+ (q^{3}+1+q^{-3}) \big( |2200 \rangle + |0220 \rangle + |0022 \rangle \big) \\
&+ [3]^{2} \big(|2020 \rangle + |2002 \rangle + |0202 \rangle \big) + (q+q^{-1})[2]^{2} \big( |1101 \rangle + |1011 \rangle \big) \\
&+ (q^{\frac{3}{2}} + q^{-\frac{3}{2}})[2][3] \big(|2101\rangle + |1201 \rangle + |1021 \rangle + |1012 \rangle \big) \\
&+ (q+q^{-1})[2][3] \big( |1102\rangle+|2011\rangle \big) + (q^{3}+1+q^{-3})[2][3] \big( |2201 \rangle + |1022 \rangle \big) \\
&+ (q^{\frac{3}{2}}+q^{-\frac{3}{2}})[3]^{2} \big( |1202 \rangle + |2102 \rangle + |2012 \rangle + |2021 \rangle \big) \\
&+ (q^{3}+1+q^{-3})[3]^{2} \big( |2202 \rangle + |2022 \rangle \big) + (q+q^{-1})^{2}[2] \big( |0111\rangle+|1110\rangle \big) \\
&+ (q^{\frac{3}{2}}+q^{-\frac{3}{2}})^{2}[3] \big( |0121\rangle+|1210\rangle \big) \\
&+ (q+q^{-1})(q^{\frac{3}{2}}+q^{-\frac{3}{2}})[3] \big(|0112\rangle+|1120\rangle+|0211\rangle+|2110\rangle \big)  \\
&+ (q^{3}+1+q^{-3})(q^{\frac{3}{2}}+q^{-\frac{3}{2}})[3] \big( |0122\rangle + |1220\rangle + |0221\rangle + |2210\rangle \big) \\
&+ (q-1+q^{-1})(q^{\frac{3}{2}}+q^{-\frac{3}{2}})[3]^{2} \big( |0212 \rangle + |2120 \rangle \big) \\
&+ (q^{3}+1+q^{-3})^{2}[3] \big( |0222\rangle + |2220\rangle \big) + (q+q^{-1})^{3}[2] |1111\rangle \\
&+ (q+q^{-1})^{2}(q^{\frac{3}{2}}+q^{-\frac{3}{2}})[3] \big( |2111\rangle + |1112\rangle \big) \\
&+ (q+q^{-1})(q^{\frac{3}{2}}+q^{-\frac{3}{2}})^{2}[3] \big( |1211\rangle + |1121\rangle \big) \\
&+ (q+q^{-1})(q^{\frac{3}{2}}+q^{-\frac{3}{2}})(q^{3}+1+q^{-3})[3] \big( |2211\rangle+|1122\rangle \big) \\
&+ (q^{\frac{3}{2}}+q^{-\frac{3}{2}})^{2}[3]^{2}(q-1+q^{-1}) \big( |2121\rangle + |1212\rangle \big) \\
&+ [3]^{2}(q-1+q^{-1})(q+q^{-1})(q^{\frac{3}{2}}+q^{-\frac{3}{2}}) |2112\rangle \\
&+ (q^{\frac{3}{2}}+q^{-\frac{3}{2}})^{2}(q^{3}+1+q^{-3})[3] |1221\rangle \\
&+ (q^{3}+1+q^{-3})^{2}[3](q^{\frac{3}{2}}+q^{-\frac{3}{2}}) \big( |1222\rangle + |2221\rangle \big) \\
&+ (q^{3}+1+q^{-3})[3]^{2} (q-1+q^{-1})(q^{\frac{3}{2}}+q^{-\frac{3}{2}}) \big( |2122\rangle + |2212 \rangle \big) \\
&+ (q^{3}+1+q^{-3})^{3}|2222\rangle
\end{align*}

\newpage

\bibliographystyle{JHEP_TD}
\bibliography{draft}

\providecommand{\href}[2]{#2}\begingroup\raggedright\begin{thebibliography}{10}

\bibitem{Ryu:2006bv}
S.~Ryu and T.~Takayanagi, {\it {Holographic derivation of entanglement entropy
  from AdS/CFT}},  {\em Phys. Rev. Lett.} {\bf 96} (2006) 181602,
  [\href{http://xxx.lanl.gov/abs/hep-th/0603001}{{\tt hep-th/0603001}}].

\bibitem{Laflorencie:2015eck}
N.~Laflorencie, {\it {Quantum entanglement in condensed matter systems}},  {\em
  Phys. Rept.} {\bf 646} (2016) 1--59,
  [\href{http://xxx.lanl.gov/abs/1512.0338}{{\tt arXiv:1512.0338}}].

\bibitem{2007JSMTE..08...24H}
M.~B. {Hastings}, {\it {An area law for one-dimensional quantum systems}},
  {\em Journal of Statistical Mechanics: Theory and Experiment} {\bf 8} (Aug.,
  2007) 08024, [\href{http://xxx.lanl.gov/abs/0705.2024}{{\tt
  arXiv:0705.2024}}].

\bibitem{Kitaev:2005dm}
A.~Kitaev and J.~Preskill, {\it {Topological entanglement entropy}},  {\em
  Phys. Rev. Lett.} {\bf 96} (2006) 110404,
  [\href{http://xxx.lanl.gov/abs/hep-th/0510092}{{\tt hep-th/0510092}}].

\bibitem{Witten89}
E.~Witten, {\it Quantum field theory and the Jones polynomial},  {\em Comm.
  Math. Phys.} {\bf 121} (1989), no.~3 351--399.

\bibitem{BFLP}
V.~Balasubramanian, J.~R. Fliss, R.~G. Leigh, and O.~Parrikar, {\it
  {Multi-Boundary Entanglement in Chern-Simons Theory and Link Invariants}},
  \href{http://xxx.lanl.gov/abs/1611.0546}{{\tt arXiv:1611.0546}}.

\bibitem{Salton:2016qpp}
G.~Salton, B.~Swingle, and M.~Walter, {\it {Entanglement from Topology in
  Chern-Simons Theory}},  {\em Phys. Rev.} {\bf D95} (2017), no.~10 105007,
  [\href{http://xxx.lanl.gov/abs/1611.0151}{{\tt arXiv:1611.0151}}].

\bibitem{RoseTub}
D.~E.~V. {Rose} and D.~{Tubbenhauer}, {\it {Symmetric webs, Jones-Wenzl
  recursions and $q$-Howe duality}},  {\em ArXiv e-prints} (Jan., 2015)
  [\href{http://xxx.lanl.gov/abs/1501.0091}{{\tt arXiv:1501.0091}}].

\bibitem{Masbaum}
G.~{Masbaum}, {\it {Skein-theoretical derivation of some formulas of Habiro}},
  {\em ArXiv Mathematics e-prints} (June, 2003)
  [\href{http://xxx.lanl.gov/abs/math/0306345}{{\tt math/0306345}}].

\bibitem{MasbaumVogel}
G.~Masbaum and P.~Vogel, {\it $3$-valent graphs and the Kauffman bracket.},
  {\em Pacific J. Math.} {\bf 164} (1994), no.~2 361--381.

\bibitem{CGV}
F.~{Costantino}, F.~{Gu{\'e}ritaud}, and R.~{van der Veen}, {\it {On the volume
  conjecture for polyhedra}},  {\em ArXiv e-prints} (Mar., 2014)
  [\href{http://xxx.lanl.gov/abs/1403.2347}{{\tt arXiv:1403.2347}}].

\bibitem{Witten89wf}
E.~Witten, {\it {Gauge Theories and Integrable Lattice Models}},  {\em Nucl.
  Phys.} {\bf B322} (1989) 629--697.

\bibitem{Witten89rw}
E.~Witten, {\it {Gauge Theories, Vertex Models and Quantum Groups}},  {\em
  Nucl. Phys.} {\bf B330} (1990) 285--346.

\bibitem{MOY}
H.~{Murakami}, T.~{Ohtsuki}, and S.~{Yamada}, {\it {Homfly polynomial via an
  invariant of colored plane graphs}},  {\em Enseign. Math.} {\bf 44} (1998)
  325--360.

\bibitem{CKM}
S.~{Cautis}, J.~{Kamnitzer}, and S.~{Morrison}, {\it {Webs and quantum skew
  Howe duality}},  {\em ArXiv e-prints} (Oct., 2012)
  [\href{http://xxx.lanl.gov/abs/1210.6437}{{\tt arXiv:1210.6437}}].

\bibitem{CGR}
S.~Chun, S.~Gukov, and D.~Roggenkamp, {\it {Junctions of surface operators and
  categorification of quantum groups}},
  \href{http://xxx.lanl.gov/abs/1507.0631}{{\tt arXiv:1507.0631}}.

\bibitem{ChunRefinedCS}
S.~Chun, {\it {Junctions of refined Wilson lines and one-parameter deformation
  of quantum groups}},  \href{http://xxx.lanl.gov/abs/1701.0351}{{\tt
  arXiv:1701.0351}}.

\bibitem{Kh}
M.~{Khovanov}, {\it {A categorification of the Jones polynomial}},  {\em ArXiv
  Mathematics e-prints} (Aug., 1999)
  [\href{http://xxx.lanl.gov/abs/math/9908171}{{\tt math/9908171}}].

\bibitem{KhR04}
M.~{Khovanov} and L.~{Rozansky}, {\it {Matrix factorizations and link
  homology}},  {\em ArXiv Mathematics e-prints} (Jan., 2004)
  [\href{http://xxx.lanl.gov/abs/math/0401268}{{\tt math/0401268}}].

\bibitem{KhR05}
M.~{Khovanov} and L.~{Rozansky}, {\it {Matrix factorizations and link homology
  II}},  {\em ArXiv Mathematics e-prints} (May, 2005)
  [\href{http://xxx.lanl.gov/abs/math/0505056}{{\tt math/0505056}}].

\bibitem{BN}
D.~{Bar-Natan}, {\it {Khovanov's homology for tangles and cobordisms}},  {\em
  ArXiv Mathematics e-prints} (Oct., 2004)
  [\href{http://xxx.lanl.gov/abs/math/0410495}{{\tt math/0410495}}].

\bibitem{Blanchet}
C.~{Blanchet}, {\it {An oriented model for Khovanov homology}},  {\em ArXiv
  e-prints} (May, 2014) [\href{http://xxx.lanl.gov/abs/1405.7246}{{\tt
  arXiv:1405.7246}}].

\bibitem{CMW}
D.~{Clark}, S.~{Morrison}, and K.~{Walker}, {\it {Fixing the functoriality of
  Khovanov homology}},  {\em ArXiv Mathematics e-prints} (Jan., 2007)
  [\href{http://xxx.lanl.gov/abs/math/0701339}{{\tt math/0701339}}].

\bibitem{Kh03}
M.~{Khovanov}, {\it {sl(3) link homology}},  {\em ArXiv Mathematics e-prints}
  (Apr., 2003) [\href{http://xxx.lanl.gov/abs/math/0304375}{{\tt
  math/0304375}}].

\bibitem{Kup}
G.~{Kuperberg}, {\it {Spiders for rank 2 Lie algebras}},  in {\em eprint
  arXiv:q-alg/9712003}, Nov., 1997.

\bibitem{MV}
M.~{Mackaay} and P.~{Vaz}, {\it {The universal sl3-link homology}},  {\em ArXiv
  Mathematics e-prints} (Mar., 2006)
  [\href{http://xxx.lanl.gov/abs/math/0603307}{{\tt math/0603307}}].

\bibitem{SN}
S.~{Morrison} and A.~{Nieh}, {\it {On Khovanov's cobordism theory for su(3)
  knot homology}},  {\em ArXiv Mathematics e-prints} (Dec., 2006)
  [\href{http://xxx.lanl.gov/abs/math/0612754}{{\tt math/0612754}}].

\bibitem{LQR}
A.~D. {Lauda}, H.~{Queffelec}, and D.~E.~V. {Rose}, {\it {Khovanov homology is
  a skew Howe 2-representation of categorified quantum sl(m)}},  {\em ArXiv
  e-prints} (Dec., 2012) [\href{http://xxx.lanl.gov/abs/1212.6076}{{\tt
  arXiv:1212.6076}}].

\bibitem{QR}
H.~{Queffelec} and D.~E.~V. {Rose}, {\it {The $\mathfrak{sl}\_n$ foam
  2-category: a combinatorial formulation of Khovanov-Rozansky homology via
  categorical skew Howe duality}},  {\em ArXiv e-prints} (May, 2014)
  [\href{http://xxx.lanl.gov/abs/1405.5920}{{\tt arXiv:1405.5920}}].

\end{thebibliography}\endgroup

\end{document}